# Effect of temperature and copper doping on the heterogeneous Fenton-like activity of $Cu_xFe_{3-x}O_4$ nanoparticles

Nahuel Nuñez[1,2,3*], Enio Lima Jr.[1,2], Marcelo Vásquez Mansilla[1,2], Gerardo F. Goya[4,5], Álvaro Gallo-Cordova[6], María del Puerto Morales[6], Elin L. Winkler[1,2,3*]

[1] Laboratorio de Resonancias Magnéticas, Gerencia de Física, Centro Atómico Bariloche, Av. Bustillo 9500, (8400) S. C. de Bariloche (RN), Argentina.

[2] Instituto de Nanociencia y Nanotecnología (CNEA-CONICET), Nodo Bariloche, Av. Bustillo 9500, (8400) S. C. de Bariloche (RN), Argentina.

[3] Instituto Balseiro, CNEA-UNCuyo, Av. Bustillo 9500, (8400) S. C. de Bariloche (RN), Argentina

[4] Dept. Física de la Materia Condensada, Universidad de Zaragoza, C/ Pedro Cerbuna 12, 50009, Zaragoza, Spain

[5] Instituto de Nanociencia y Materiales de Aragón, CSIC-Universidad de Zaragoza, C/ Mariano Esquillos S/N, 50018, Zaragoza, Spain

[6] Instituto de Ciencia de Materiales de Madrid, ICMM/CSIC, C/ Sor Juana Inés de la Cruz 3, 28049, Madrid, Spain

*Corresponding authors : nahuel.nunez@ib.edu.ar

## Abstract

Ferrite nanoparticles serve as potent heterogeneous Fenton-like catalysts, producing reactive oxygen species (ROS) for decomposing organic pollutants. We investigated the impact of temperature and copper content on the catalytic activity of nanoparticles with different oxidation states of iron. Via solvothermal synthesis, we fabricated copper-doped magnetite ($Cu_xFe_{3-x}O_4$) with a $Fe^{2+}$/Fe ratio ~0.33 for the undoped system. Using a microwave-assisted method, we produced copper-doped oxidized ferrites, yielding a $Fe^{2+}$/Fe ratio of ~0.11 for the undoped nanoparticles. The ROS generated by the catalyst were identified and quantified by electron paramagnetic resonance, while optical spectroscopy allowed us to evaluate its effectiveness for the degradation of a model organic dye. At room temperature, the magnetite nanoparticles exhibited the most •OH radical production and achieved almost 90% dye discoloration in 2 hours. This efficiency decreased with increasing Cu concentration, concurrently with a decrease in •OH generation. Conversely, above room temperature, Cu-doped nanoparticles significantly enhance the dye degradation, reaching 100% discoloration at 90°C. This enhancement is accompanied by a systematic increase in the kinetic constants, obtained from reaction equations, with Cu doping. This study highlights the superior stability and high-temperature catalytic advantages of copper ferrite holding promise for enhancing the performance of nanocatalysts for decomposing organic contaminants.

# 1. Introduction

Environmental remediation processes based on the catalytic activity of magnetic nanoparticles have gained considerable attention in recent years.[1–3] One promising example on this is the use of iron oxide nanoparticles as heterogeneous catalysts in the Fenton reactions that generates reactive oxygen species (ROS).[4,5] These species allows the degradation of various organic pollutants through an advanced oxidation process.[6] In the homogeneous Fenton reaction, •OH and •OOH radicals are generated by soluble $Fe^{2+}$ and $Fe^{3+}$ ions through the following reactions: $Fe^{2+} + H_2O_2 \rightarrow Fe^{3+} + \bullet OH + OH^- \; (k_{cat} = 63 \; M^{-1} \; s^{-1})$ and

$Fe^{3+} + H_2O_2 \rightarrow Fe^{2+} + \bullet OOH + H^+ \; (k_{cat} = 0.001 \; M^{-1} \; s^{-1})$.[7] In the literature this second equation is known as Fenton-like reaction, being the limiting reaction due to the lower rate constant.[8,9] These free radicals oxidize organic molecules and pollutants, reducing their toxicity. Compared to the homogeneous Fenton reaction, the use of heterogeneous catalysts offers several advantages, such as increased stability, recyclability, and reduced secondary pollution.[10] It is expected that in homogeneous catalysis, the solubility of catalyst ions is the process limiting factor, while in the heterogeneous one, the adsorption steps as well as the diffusion of the reactive species in the medium are dominant.[11] Therefore, heterogeneous catalysis provides greater degrees of freedom to tune and gain control over the subsequent advanced oxidation reaction.

Like any thermally activated process, the kinetics of the Fenton reaction is accelerated by temperature.[12] For this reason, numerous studies aimed to design multifunctional nanoparticles capable of inducing local heating to enhance its catalytic activity.[13,14] This provides an additional parameter to control and optimize the efficiency of catalysts in Fenton processes. In fact, studies focused on decontaminating landfill leachate using copper catalysts concluded that the optimum temperature for the reaction was 70 °C.[15]

While it is true that increasing the temperature of the system can elevate the cost of wastewater treatment, in some instances, the waste is already above room temperature, condition that can be leveraged to enhance the efficiency of the nanocatalyst. For example, it was reported that the temperature of effluents from textile wet processing is in the range of 30°C to 60°C.[16] Similarly, vinasse, a byproduct of bioethanol distillation, is generated at high temperatures, approximately around 90°C.[17,18] Another example is the paper industry, where the chemical treatment of wood generates a black liquor residue at about 60-70°C.[19] On the other hand, while increasing the temperature accelerates the kinetics of the reaction, improving the catalyst's efficiency, it may also produce undesirable effects The thermally accelerated decomposition of hydrogen peroxide can decrease the degradation efficiency by Fenton and Fenton-like processes.[20] Moreover, it has been demonstrated that $Fe_3O_4$ nanoparticles undergo a transformation into $\gamma$-$Fe_2O_3$ after continuous Fenton reaction due to the different $k_{cat}$ of $Fe^{2+}$ and $Fe^{3+}$ ions, resulting in a subsequent loss of their catalytic activity.[21] This oxidation of iron catalysts also could be promoted by increasing the temperature. Therefore, various effects of the operating temperature must be taken into account when the optimal working temperature for wastewater treatment is defined.

To optimize the catalytic efficiency of iron oxide nanoparticles in advanced oxidation reactions, various ions have been incorporated into the ferrite lattice in order to modify their surface reactivity.[7,22,23] Among them, copper has received special attention due to its strikingly similar redox properties like iron.[24–32] Both the monovalent $Cu^+$ and divalent $Cu^{2+}$ oxidation states easily react with $H_2O_2$ analogous to the $Fe^{2+}/H_2O_2$ and $Fe^{3+}/H_2O_2$ systems, the reaction constants of these Fenton-like processes are $k_{cat} = 10^4\ M^{-1}\ s^{-1}$ and $k_{cat} = 4.6 * 10^2\ M^{-1}\ s^{-1}$, respectively.[7] Due to its enhanced activity, copper nanocatalysts have been previously employed to efficiently degrade contaminants like phenols,[33] insecticides,[34] pharmaceuticals[35–37] and also as antibacterial agent.[38] Moreover, a synergistic effect was observed when copper was introduced into a mesoporous iron oxide catalyst for the Fenton-like process.[15,26,39,40] In this case, the proposed mechanism for the good performance of copper-iron catalysts is the regeneration of $Fe^{2+}$ ions through the reaction $Fe^{3+} + Cu^{1+} \rightarrow Fe^{2+} + Cu^{2+}$. Another interesting capability of copper catalysts is its good performance after recycling. Hussain et al. found that zirconia-supported copper catalysts maintained their activity when recycled, unlike iron catalysts whose activity diminished.[40] For these reasons copper catalysts are interesting materials to be tested in Fenton-like processes at high temperature reactions.

In this context, in the present work we investigate the heterogeneous Fenton-like catalytic activity of copper-doped iron oxide nanoparticles with the aim of studying the role played by the surface active ions in the generation of ROS, and determine the proper doping condition and surface oxidation state for different working temperature in wastewater treatment. For this, copper-doped magnetite and maghemite nanoparticles were synthesized by two different polyol methods: solvothermal and microwave-assisted, respectively. The as-synthesized nanoparticles were characterized by various techniques, including X-ray diffraction (XRD), transmission electron microscopy (TEM) and X-ray photoelectron spectroscopy (XPS) in order to determine their crystalline structure, morphology and composition. Subsequently, the production of free radicals catalyzed by each sample was identified and quantified by means of electron paramagnetic resonance (EPR) spectroscopy assisted with a spin trap molecule. Finally, the catalytic performance of the nanoparticles was evaluated by measuring their degradation efficiency of methylene blue (MB) at different reaction temperatures using optical spectroscopy. These results demonstrate that, while the magnetite is most efficient nanocatalyst at room temperature, it oxidizes and loses its activity at higher temperatures; copper doping being essential to maintain, and even surpass, its performance at higher temperatures.

## 2. Materials and methods

### 2.1. Materials

The reagents used in this work are: iron (III) nitrate nonahydrate ($Fe(NO_3)_3 . 9H_2O$, 98% Sigma-Aldrich), copper (II) sulfate pentahydrate ($CuSO_4 . 5H_2O$, 99% Merk), triethylene glycol (99% Sigma-Aldrich), diethylene glycol (99% Sigma-Aldrich), ethanol (96%), methylene blue (>82% Sigma-Aldrich), 5,5-dimethyl-1-pyrroline N-oxide (DMPO, >97% Cayman), dimethyl sulfoxide (DMSO, >99%) and $H_2O_2$ aqueous solution (30% Sigma-Aldrich).

## 2.2.    Synthesis of the catalysts

For this study we synthesized two series of copper doped iron oxide nanoparticles. The first one is composed of almost stoichiometric copper ferrite nanoparticles ($Cu_x^{2+}Fe_{1-x}^{2+}Fe_2^{3+}O_4$). It was obtained by the solvothermal method using a polyol as solvent. The second batch was obtained by the microwave assisted method, also using a polyol as solvent. In this case the nanoparticles were overoxidized by adding water into the synthesis reactor.

Solvothermal synthesis: In this method 4 mmol of metals were dissolved in 40 mL of triethylene glycol. The proportions of the precursors $Fe(NO_3)_3 \cdot 9H_2O$ and $CuSO_4 \cdot 5H_2O$ were adjusted according to the required stoichiometry by the relations: $m_{Fe(NO_3)_3} = (3-x) * 241.86$ mg and $m_{CuSO_4} = x * 249.68$ mg, where x determines the Cu content of atoms in a molecule ($Cu_xFe_{3-x}O_4$). The solution was heated at 100 °C for 1 h to evaporate the water from the system and then transferred to a 100 mL teflon autoclave and kept at 260 °C for 4 h. Once the synthesis is finished, the obtained material was repeatedly washed by magnetic separation with ethanol and then dried at 70 °C. The nanoparticle samples were named STX, where X=0, 1, 2 and 3 represents the nominal concentration x=0, 0.1, 0.2 and 0.3, respectively.

Microwave-assisted synthesis: In this synthesis method, the microwave oven Monowave 300 (Anton Paar GmbH, Graz, Austria) with a built-in magnetic stirrer was used to synthesize the catalysts, working at 2.45 GHz and following a protocol similar to that presented by Gallo-Cordova et al[1]. Briefly, 1.75 mmol of metals were dissolved in a solution of diethylene glycol (18.3 mL) and water (0.7 mL), adjusting the proportions of precursors $Fe(NO_3)_3 \cdot 9H_2O$ and $CuSO_4 \cdot 5H_2O_2$ according to the required stoichiometry following the relationship previously mentioned in the solvothermal synthesis method. The solution was transferred to a 30 mL vial and the temperature was raised to 230 °C with a ramp of 5.25 °C/min, where it was maintained for 2 h and then cooled abruptly. It is important to mention that the pressure increases due to the presence of water that lowers the boiling point of the solvent mixture (170 ºC), reaching up to 30 Bar in some cases. After the synthesis, the nanoparticles were repeatedly washed by magnetic separation with alcohol and redispersed in water. In this case, the nanoparticle samples were named MWX, where X=0, 1, 2, 3 and 4 represents the nominal concentration x=0, 0.1, 0.2, 0.3 and 0.4, respectively.

## 2.3. Characterization of the samples and evaluation of catalytic activity

The crystal structure of the synthesized samples was characterized by X-ray diffraction (XRD) using the Bruker Advance D8 diffractometer (Cu-$\mathrm{K}_\alpha$ radiation, $\lambda$=0.15406 nm). The incorporation of copper ions into the structure of the nanoparticles was verified by inductively coupled plasma optical emission spectroscopy (ICP-OES) using a Perkin Elmer apparatus (OPTIME 2100 DV) and the organic content quantified by thermogravimetric analysis (TGA) in a ATD/DSC/TG, Q600 from TA Instruments. X-ray photoelectron spectroscopy (XPS) was used to analyze the oxidation state of iron ions in the nanoparticles, employing the Kratos AXIS Supra. The values of $Fe^{2+}$ and $Fe^{3+}$ for each sample were obtained by fitting each XPS spectrum with the peaks of the $Fe^{2+}$ (lower binding energy) and $Fe^{3+}$ (higher binding energy) multiplets using the software CASAXPS. The size and morphology of the NPs were studied by transmission electron microscopy (TEM) in a Philips CM-200 microscope operating at 200 kV. The DC magnetization of the samples was studied with a LakeShore 7300 vibrating sample magnetometer (VSM). Magnetization versus applied field (M(H)) cycles were acquired at room temperature with the VSM up to an applied field of ±10 kOe.

The generation of free radicals by the catalysts was studied by electron paramagnetic resonance (EPR) working in the X-band (9.5 GHz) at room temperature with a BRUKER ELEXSYS II-E500 spectrometer using the nitrone-based DMPO as a spin trap. Measurements were performed with a modulation signal of 100 kHz and 3 G of amplitude, and using the resonance of $Mn^{2+}$ impurities in a MgO crystal as a pattern signal to normalize the free radical production of each sample.[21] In these experiments, 0.1 mg/mL of catalyst was dispersed in 100 mM acetate buffer at pH=5, then 50 μL of a 1 mg:6 mL solution of DMPO:DMSO were added. Then 5 μL of 30% $H_2O_2$ aqueous solution were added and EPR spectra were taken at intervals of no more than 10 minutes. About 90 μL of the reaction solution was contained in a quartz tube and the height of the measured region was 30 mm.

To evaluate the efficiency of the synthesized nanoparticles for degrading organic compounds, colorimetric experiments were performed with methylene blue dye. The methylene blue concentration was selected from previous works[42], and the catalyst and hydrogen peroxide dosage were fixed in common values reported for Fenton and Fenton-like processes[43,44]. A pH=5 was chosen, acidic enough to favor the Fenton reaction[45], but not so low to avoid leaching, which is drastically enhanced for pH<5.[46,47]

In these experiments, 1 mg/mL of nanoparticles and 100 ppm of dye were dispersed in a pH=5 100 mM acetate buffer, and the system was kept under agitation for two hours to ensure adsorption of the dye onto the nanoparticles' surface. Then, 10 µL/mL of 30% $H_2O_2$ aqueous mixture were added to the solution, and the degradation efficiency was determined by measuring the absorbance at 663 nm at different time intervals (5, 15, 30, 60 and 120 min) and considering a calibration curve absorbance vs MB concentration. These experiments were carried out in 2 mL reactors, using different reactors for each time (six in total for a complete measurement). Measurements were taken at room temperature, at 60 °C and at 90°C using the NUMAK 721 UV-Vis spectrophotometer.

## 3. Results and discussion

### 3.1. Nanocatalysts characterization

XRD patterns of the copper doped iron oxide nanoparticles synthesized by the solvothermal and microwave-assisted method are shown in **Fig.1a** and **Fig.1b**, respectively. The main XRD diffraction peaks observed can be assigned to the Fd3m spinel structure, characteristic of copper ferrite. Noticeably, the XRD pattern of the x=0 microwave-assisted sample shows also additional peaks at 32° and 34°, besides the ones indexed with the spinel structure mentioned above, which are characteristic of the ordering of cation vacancies in the maghemite structure.[48] This result confirms a higher degree of oxidation of the samples fabricated by microwave-assisted method in comparison to those prepared by the solvothermal one, attributed to the presence of the water added in the synthesis. The intensity of the mentioned peaks decreases as copper is introduced into the structure and almost disappears for x=0.2, which could be explained by the copper occupancy of cation vacancies into the maghemite structure. Minority metallic copper segregation was observed for the largest copper substitution, this occurs for x>0.3 and x>0.4 for nanoparticles fabricated by solvothermal and microwave-assisted methods, respectively. Due to the metallic phase segregation, this study was restricted to x≦0.3 and x≦0.4 for nanoparticles synthesized by solvothermal and microwave route, respectively. The XRD patterns of the solvothermal samples show a low intensity peak at 24°, attributed to the presence of iron hydroxide traces (JCPDS 00-046-1436) [49]. Additionally, narrow peaks at 17° and 23° were observed in the x=0 sample which are ascribed to polymerized (poly-) ethylene glycol.[50,51]

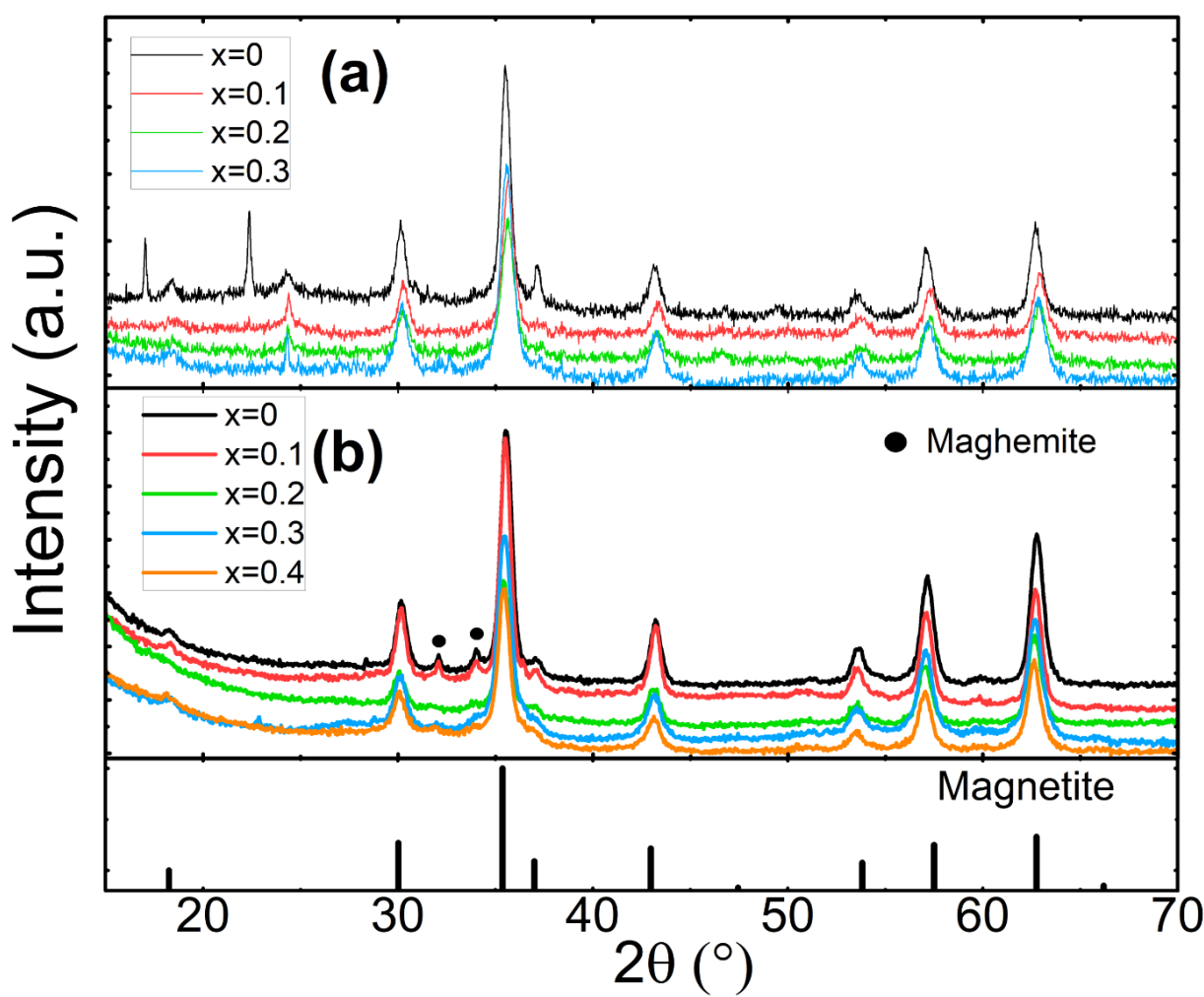

**Fig.1:** X-ray diffractogram of the copper-incorporated iron oxide nanoparticles obtained by: **(a)** the solvothermal method and **(b)** the microwave–assisted method.

The incorporation of copper in the structure was measured by ICP-OES, and the results are presented in **Table1**, where a systematic increase of Cu-concentration was determined, although slightly smaller than the nominal doping for both methods. The oxidation state of iron for all the nanoparticles systems was evaluated by XPS measurements. Notice that although the XPS is a surface sensitive technique, in this case it provides information of almost the whole nanoparticle as the measuring range technique is about 4 nm. From the XPS spectra (shown in **Figs. S1** and **S2** of the Supplementary Information) the $Fe^{2+}/Fe$ ratio was determined for all the samples and the results are presented in **Table1**. In the solvothermal samples, the $Fe^{2+}/Fe$ ratio obtained by XPS is similar to the one expected for the stoichiometric $Cu_xFe_{3-x}O_4$ nanoparticles, where x corresponds to the value determined by ICP-OES. On the other hand, microwave-assisted samples present much lower $Fe^{2+}/Fe$ ratio than the theoretical one, confirming that these samples are overoxidized. This result is consistent with the maghemite phase detected by the XRD patterns.

| Method | Solvothermal | | | | Microwave-assisted | | | | |
|---|---|---|---|---|---|---|---|---|---|
| Sample | ST0 | ST1 | ST2 | ST3 | MW0 | MW1 | MW2 | MW3 | MW4 |
| $x_{nominal}$ | 0 | 0.1 | 0.2 | 0.3 | 0 | 0.1 | 0.2 | 0.3 | 0.4 |
| $x_{ICP}$ | 0 | 0.08 | 0.15 | 0.23 | 0 | 0.09 | 0.17 | 0.24 | 0.32 |
| $Fe^{2+}/Fe$ nominal | 0.333 | 0.315 | - | 0.278 | 0.333 | 0.313 | 0.293 | 0.286 | 0.253 |
| $Fe^{2+}/Fe$ by XPS | 0.33 | 0.32 | - | 0.26 | 0.11 | 0.16 | 0.18 | 0.12 | 0.16 |
| $d_{TEM}$ (nm) | 13(3) | 12(3) | 10(2) | 9(2) | 11(2) | 12(2) | 11(2) | 12(4) | 11(3) |
| $(\frac{m_{NPs}}{m_{sample}})_{TGA}$ | 0.81 | 0.84 | 0.83 | 0.86 | 0.96 | 0.97 | 0.96 | 0.96 | 0.95 |

| $M_s$ by VSM (emu/g) | 66(1) | 67(1) | 64(1) | 62(1) | 76(1) | 74(1) | 68(1) | 51(1) | 50(1) |
|---|---|---|---|---|---|---|---|---|---|

**Table1:** Results of the $Cu_xFe_{3-x}O_4$ nanoparticles characterization: Cu incorporation degree determined by ICP-OES ($x_{ICP}$), $Fe^{2+}/Fe$ ratio determined by XPS measurements and theoretically by taking in account the copper incorporation obtained by ICP-OES, size obtained by TEM ($d_{TEM}$), mass percentage of nanoparticles in the samples as obtained by TGA ($\frac{m_{NPs}}{m_{sample}}$), and the saturation magnetization ($M_s$) obtained from the M(H) curves at room temperature.

To study the size and morphology of the nanoparticles we performed TEM measurements of all samples, which are presented in **Fig.2** along with the corresponding size distribution histogram (measured as diameter due to the spherical-like shape of the particles) and fitted to a lognormal function. Solvothermal samples exhibit a slight dependence of size with copper content, ranging from 13(3) nm for x=0 to 9(2) nm for x=0.3 and irregular shape (see **Table1**). In contrast, no size dependence with copper content was observed in microwave-assisted samples, all with rather spherical shape and sizes between 11 - 12 nm, probably due to the higher pressure synthesis condition.

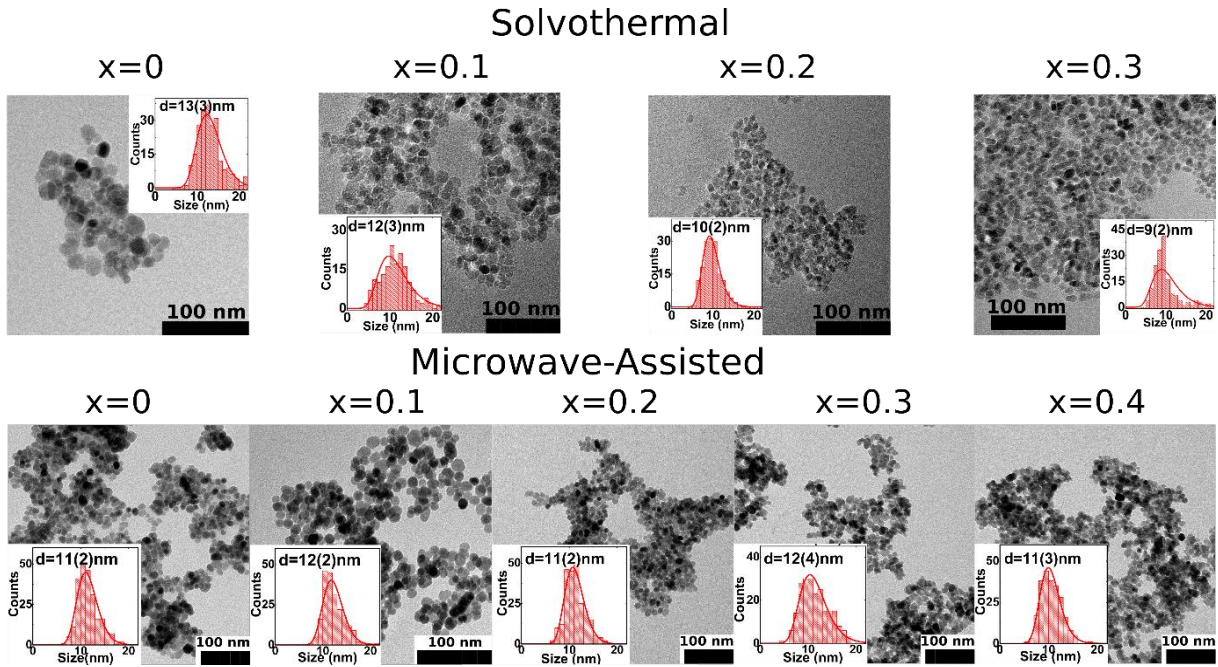

**Fig.2:** Representative TEM images of the MWX and STX $Cu_xFe_{3-x}O_4$ nanoparticles and their corresponding size distribution histogram fitted with lognormal distribution.

The DC field dependence of the magnetization, measured at room temperature, shows a reversible behavior for all the samples, in agreement with a superparamagnetic regime of the nanoparticles at these conditions. **Figures 3a** and **3b** show the MvsH curves for the solvothermal and microwave-assisted samples, respectively. The sample´s magnetization was corrected by considering the proportion of the organic compound in the as-made nanoparticles, determined from TGA measurement, as reported in **Table1**.

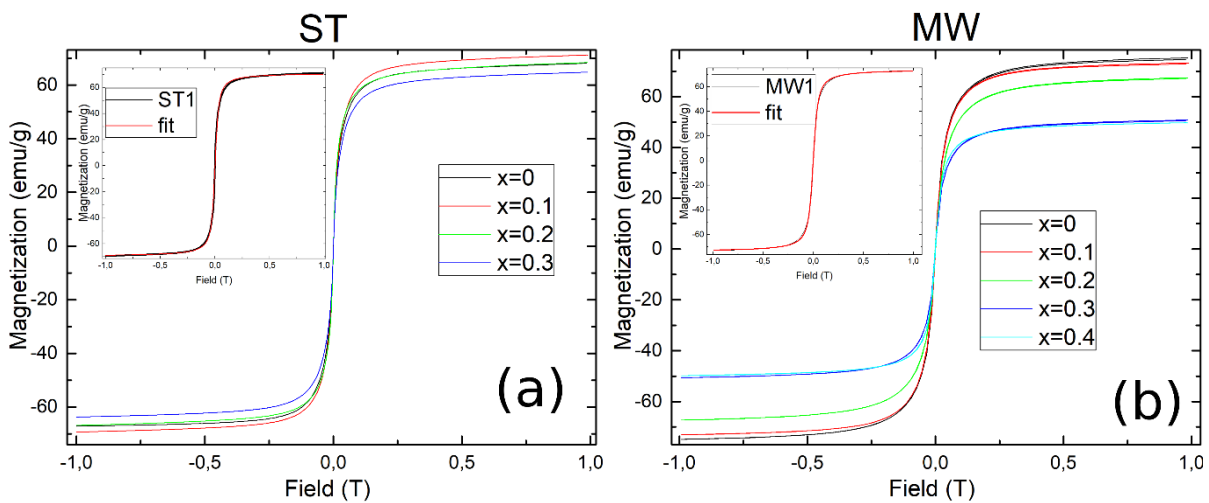

**Fig.3: (a)** M vs H cycle of the solvothermal nanoparticles. The inset shows the fit using the Langevin model for the ST1 sample. **(b)** M vs H cycle of the Microwave-assisted nanoparticles. The inset shows the fit using the Langevin model for the MW1 sample.

The M(H) curve can be fitted with a Langevin model for all the samples, in agreement with a superparamagnetic behavior. From the fitting, the corresponding saturation magnetization ($M_s$) can be determined. **Table1** presents the values of $M_s$ obtained for both the solvothermal and microwave-assisted samples. In both systems it is observed a decreases of $M_s$ when the copper concentration increases in the ferrite, as expected when the $Fe^{2+}$ ($\mu \sim 4\mu_B$) is replaced by an ion with smaller magnetic moment such as $Cu^{2+}$ ($\mu \sim 1\ \mu_B$) and in agreement with previous results.[52,53] From this results it is also noticed that the microwave-assisted nanoparticles (MW) have larger magnetization than those prepared by solvothermal method (ST). This is striking considering that the crystalline structure of MW nanoparticles corresponds to the maghemite oxidized phase with lower $M_s$ than magnetite, i.e. $M_s(\gamma\text{-}Fe_2O_3)$=74 emu/g and $M_s(Fe_3O_4)$=84 emu/g for bulk materials.[54] The origin of this result may be due to the higher pressure attained in the microwave synthesis that improve the crystallinity of the nanoparticles, reducing magnetic disorder, and also due to the overestimation of the magnetite mass because of the presence of iron hydroxide impurities in solvothermal samples. At this point we would like to highlight that these systems show promising properties to test their potential as magnetic catalysts. On one hand, the nanoparticles' superparamagnetic behavior at room temperature reduces the agglomeration and facilitates the nanoparticles dispersion in the solution; and on the other hand, their relative large magnetization, i.e. $M_s$=50-70 emu/g, would enable the nanoparticles magnetically-assisted harvesting from solution after catalytic reactions, both properties particularly relevant for the design of magnetic nanocatalysts for environmental remediation applications.

### 3.2. Catalytic evaluation

In order to evaluate the Fenton-like activity of these nanocatalysts, we performed EPR measurements of the nanoparticles dispersed in acetate buffer solution at pH=5, containing $H_2O_2$ and using DMPO as spin trap. These measurements allow the identification and quantification of the free radicals produced in the reaction. **Figure 4a** shows a representative EPR spectrum along with its corresponding fitting curve. The fitting curve results to the superposition of the resonance lines of the different free radicals generated in the reaction. According to the NIEHS Spin-Trap database,[55] from the spectrum features and the fitting parameters, the resonance of four different free radicals can be clearly recognized: •OH, •OOH, •CH₃, and •N. According to the Fenton-like reaction, the •OH and •OOH radicals were produced by $Fe^{2+}/Cu^{+}$ and $Fe^{3+}/Cu^{2+}$, respectively; while the •CH₃ signal comes from a secondary reaction between •OH and the DMSO used to dissolve DMPO.[23] Also, a small nitrone radical signal is observed. However, this last one is also observed in the solution without the nanoparticles, being not resulting from oxidation process during the studied reactions, so it was not considered in the analysis.[23,56] All the spectra also shows the resonance corresponding to the MgO: $Mn^{2+}$ sample used as a pattern to normalize the EPR intensity.

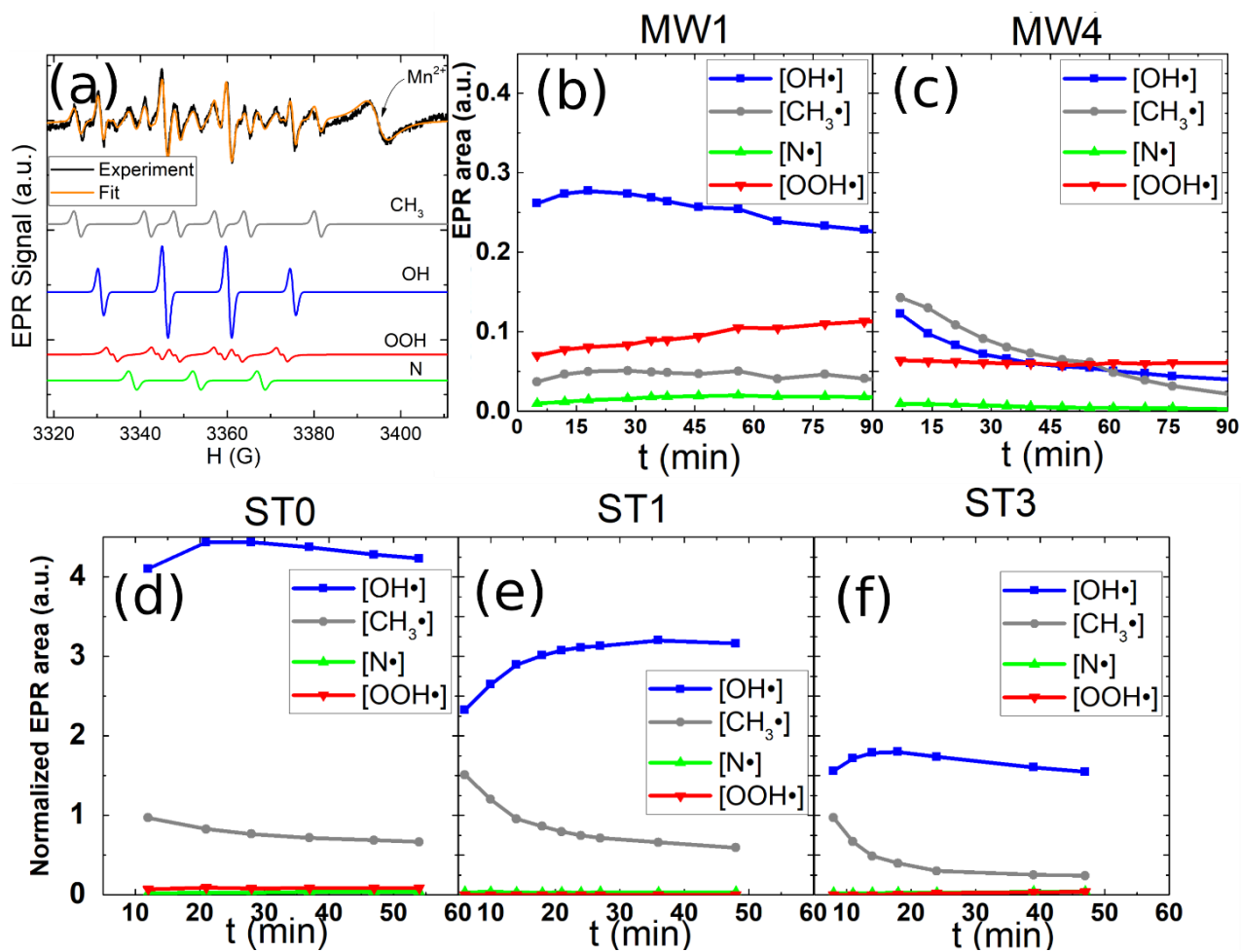

**Fig.4: (a)** Representative measurement of the free radicals generated by the catalysts, along with the spectrum fitting and the corresponding deconvoluted spectra. Kinetic curves of the free radicals catalyzed by the Cu-doped oxidized nanoparticles fabricated by microwave-assisted method **(b-c)**; and, **(d-f)** by the cu-doped magnetite nanoparticles fabricated by solvothermal method.

The kinetic reaction was followed by acquiring the EPR spectra for each set of samples as a function of the time. The quantity of radicals is directly proportional to the area of the EPR absorption curve, which can be determined by the double integral of the measured spectrum.[56] The kinetic of radical generation by the Cu-doped magnetite samples, ST0, ST1, and ST3 is shown in **Figs.4d-f**. From these figures it is notice that the •OH and •CH$_3$ are the main species produced, with a systematic decrease of their concentration for increasing copper content. Assuming that $Cu^{2+}$ replaces $Fe^{2+}$ within the spinel structure, this result suggests that $Fe^{2+}$ ions have the highest catalytic activity at room temperature. On the other hand, the Cu-doped oxidized samples, i.e. the batch fabricated by microwave assisted method; produce at least ten times fewer radicals than the Cu-doped magnetite, as depicted in **Figs.4b-c**. This result is in agreement with the lower $Fe^{2+}/Fe$ ratio in this set of samples measured by XPS, as compared with the Cu-doped magnetite. Consistently, a lower ROS concentration is expected due to the lower rate constant of $Fe^{3+}$ than that of $Fe^{2+}$ in the Fenton reaction. Notice that, besides the •OH and •CH$_3$, microwave-assisted samples also produce an appreciable amount of •OOH radicals, as shown in **Figs.4b-c**. Actually, in sample MW4 the •OOH intensity exceeds that of the hydroxyl radical after 1 h of reaction time. This response is attributed to the predominance of $Fe^{3+}$ oxidation state of iron.[13,23] Overall, the EPR measurements provide valuable insights into the mechanisms of free radicals generation by the copper ferrite nanoparticles, highlighting the importance of $Fe^{2+}$ ions and the impact of copper incorporation on the Fenton catalytic activity of the nanoparticles at room temperature.

In order to evaluate the ability of the synthesized nanoparticles to degrade organic compounds, we conducted colorimetric experiments using a MB cationic dye. In these experiments the nanocatalysts and the MB were dispersed in a pH=5 acetate buffer and, to ensure adsorption of the dye onto the nanoparticles surface, the system was kept under agitation for two hours before adding $H_2O_2$. Furthermore, control experiments, or blanks tests, were conducted for each condition to quantify MB degradation by hydrogen peroxide alone, in absence of catalyst. **Figures 5a** and **5b** show the MB discoloration curves measured at room temperature using the nanocatalyst fabricated by solvothermal and microwave-assisted methods, respectively. Nanoparticles fabricated by solvothermal route showed the great performance for degrading MB, since the magnetite ST0 sample exhibited a percentage of discoloration up to 80% in the first 15 minutes and up to 90% in 2 h. For this set of samples the activity decreases with the copper incorporation, for example the ST3 nanoparticles only exhibited an efficiency of 24% in 2 h. The activity observed at room temperature for this family of samples can be attributed to the $Fe^{2+}$ active ions at the nanoparticle surface and, accordingly, the efficiency decreased with the substitution of iron by copper. The EPR measurements support this result as the $Fe^{2+}$ was identified as the principal responsible of the free radical production at room temperature. On the other hand, none of the Cu-doped oxidized nanoparticles, fabricated by microwave-assisted method, were efficient for degrading MB at room temperature (**Fig. 5b**). Result in agreement with the low $Fe^{2+}/Fe$ ratio obtained by XPS for these samples and the low free radical production measured by EPR experiments.

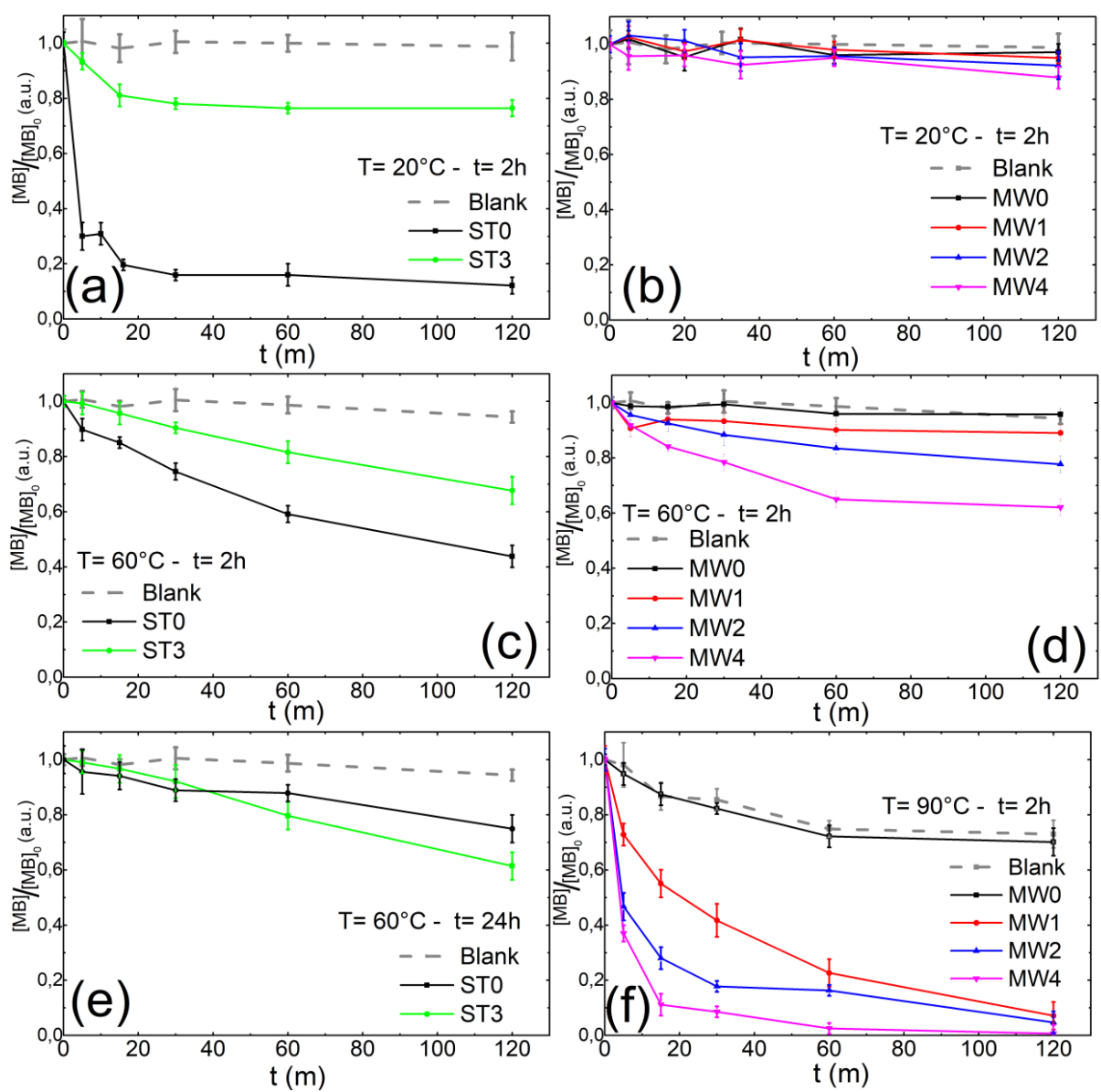

**Fig.5:** Methylene blue degradation experiments by solvothermal samples measured **(a)** at room temperature with an adsorption time of t=2h; **(c)** at T=60°C and adsorption time of t=2h; and **(e)** at T=60°C with an adsorption time of t=24h. Methylene blue degradation experiments by microwave-assisted samples measured **(b)** at room temperature with an adsorption time of t=2h; **(d)** at T=60°C

with an adsorption time of t=2h; and **(f)** at T=90°C with an adsorption time of t=2h. Control experiments, indicated by dashed lines in each graph, were performed to quantify MB degradation by hydrogen peroxide alone, without the catalysts.

It is well known that the catalytic reaction rate increases with the increasing temperature due to the higher kinetic energy of the molecules. Besides, as mentioned at the introduction, different residual effluents are produced above room temperatures, as it is the case of textile and paper industries [16–19] Therefore, it is interesting to take advantage of this additional kinetic energy present in some wastewater to increase the performance of the nanocatalysts to decompose organic contaminants. To analyze the temperature dependence on the reactions and the stability of the materials, we performed colorimetric experiments up to 90°C with a similar protocol to the one mentioned above.

**Figure 5c** shows the MB oxidation experiments at 60°C using solvothermal samples. Surprisingly, the MB degradation is lower at 60°C than the obtained at room temperature for the ST0 sample. This result can be explained by considering the accelerated oxidation from $Fe^{2+}$ to $Fe^{3+}$ as a consequence of the temperature, which decreases the kinetic rate for the ROS production. For the ST3 sample a slight increase of activity was observed at higher temperature, attributed to the combined effect of iron oxidation and kinetic promotion of Cu catalyzed degradation pathways, expected to rapidly increase with increasing temperature. This assumption was confirmed by repeating the discoloration curve after incubating the nanoparticles for 24 h at 60°C, to complete the surface oxidation (**Fig.5e**). In this case the efficiency of ST0 sample was even lower than the presented in **Figs. 5a** and **5c**, consistently with the lower kinetic of $Fe^{3+}$ to catalyze the $H_2O_2$ decomposition in ROS. On the other hand, the ST3 activity remained almost the same for both of the adsorption times tested, suggesting a better response and stability of copper doped nanoparticles to high temperature conditions.

This observation is clearly confirmed by the discoloration experiments using the Cu-doped maghemite nanoparticles (samples fabricated by microwave route) above room temperature.

**Figure 5d** shows the experiments carried out at 60°C, where it is observed that almost all the samples showed catalytic activity to degrade MB and its efficiency increased with the copper content. When the MB oxidation experiment is running at 90°C the response of the nanocatalysts improves significantly, reaching up to 90% percentage of MB discoloration in the first 30 minutes and almost 100% in 2 h. This result shows that the copper ferrite exhibits superior thermal stability and catalytic performance under elevated temperature conditions, maintaining its catalytic activity without significant degradation or loss of performance. It is noteworthy that increasing the temperature in reactions mediated by iron-based catalyst is not always beneficial, and it is crucial to know the chemical kinetic of the active ions involved in the ROS generation, for the design of proper nanocatalys for specific work conditions. The key role of copper ions in enhancing the catalytic activity of oxidized nanoparticles can be elucidated through an examination of the electronic properties of Cu-doped maghemite. Employing Density Functional Theory (DFT) alongside with the functional Perdew, Burke and Ernzerhof (PBE) [57], Pires et al. [58] calculated the electronic density map of $\gamma$-$Fe_2O_3$ and Cu/ $\gamma$-$Fe_2O_3$ at the (311) surface plane. These computations revealed a lower electronic density at Cu sites, these more positive region exhibit heightened susceptibility to interact with hydrogen peroxide molecules, thereby facilitating the generation of •OH radicals. Despite the advancements achieved through computational methods, a comprehensive investigation of the electron transfer process at interfaces is still lacking to fully understand the peroxidase decomposition mechanism at the nanocatalyst's surface.[59]

Nanocatalyst stability is also an important property desired in materials for environmental applications. With the aim to evaluate the stability of the samples with time, the free radical production was followed by EPR after different storage conditions. **Figure S3** of the Supplementary Information shows the •OH radicals produced by $Cu_{0.1}Fe_{2.9}O_4$ (MW1) and $Cu_{0.4}Fe_{2.6}O_4$ (MW4) nanoparticles upon synthesis and after 4 months of storage in air and water. The •OH production in the as-made samples decreases with the copper content due to copper substituting the most active $Fe^{2+}$ ion. For powder samples stored in air condition, $Fe^{2+}$ oxidized to

$Fe^{3+}$, decreasing MW1's activity. Conversely, MW4, already highly oxidized and copper-activity reliant, retains its •OH production after 4 months in air. This effect is amplified in water storage, where after four months, the trend reverses, and •OH production increases with copper content. These results signal that the $Fe^{2+}$-dependent catalysts would become less efficient upon reuse, while copper-dependent catalysts are expected to maintain their efficiency for longer times.

On the other hand, the stability of the catalysts before and after the MB degradation experiments, were conducted by evaluating the changes in the structure and morphology of the nanoparticles. The degradation was carried out under the most efficient conditions for each catalyst, i.e. room temperature for magnetite (ST0) and high temperature (60°C) for $Cu_{0.4}Fe_{2.6}O_4$ (MW4). **Figure S4-a** and **c**, included in the Supplementary Information, illustrate the diffractograms comparison before and after catalysis for samples ST0 and MW4, respectively. For ST0, the spinel phase characteristic peaks remain unchanged. Also it is observed that the peaks at 17° and 23° ascribed to polymerized ethylene glycol are no longer observed after reaction, likely due to the high solubility of the ethylene glycol in water. Additionally, the peak at 24°, attributed to traces of iron hydroxide, vanishes as expected, given its lower stability compared to the ferrite structure of the nanoparticles. No structural changes are observed in the MW4 sample after the reaction, according to the X-ray diffractograms. TEM micrographs of samples ST0 and MW4 after catalysis are presented in **Fig. S4-b** and **d**, respectively, along with their corresponding size distribution pre- and post-catalysis. The size distributions for ST0 and MW4 before and after catalysis show no significant differences, suggesting minimal or no leaching during the process.

The Fenton-like reaction assumes that the free radicals produced in the nanoparticles' surface react with the organic molecules adsorbed on the surface of the nanocatalyst. This oxidation reaction can be described by the nth-order equation:

$$\frac{dC}{dt} = -k_n C^n$$

where $C$ represents the organic molecules concentration, $k_n$ is the reaction constant and $n$ is the order of the reaction. Usually, the Fenton-like oxidation follows a pseudo-second order dependence $C_0/C = k_2 C_0 t + 1$ , i.e. the kinetic depends on the pollutant or dye concentrations.[60,61] On the contrary, if the reaction is independent of the substrates concentration, the kinetics follows a pseudo-first order dynamics $ln\left(\frac{C_0}{C}\right) = k_1 t$.[8,62] **Figures S5** and **S6** in the Supplementary Information include the analysis of the results, adjusted with the pseudo-first order and pseudo-second order models, respectively; and the corresponding fitting parameters are reported in **Table 2**. At room temperature, the system with higher pseudo-first order kinetic rate is the magnetite (ST0), reaching $k_1$=0.07(2) min⁻¹ value, while in the doped samples the activity is negligible. Instead, the constant rate of the MB degradation by over oxidized MWX samples at 90 °C, systematically increases with the Cu concentration up to $k_1$= 0.10(2) min⁻¹ for the sample MW4 ($Cu_{0.4}Fe_{2.6}O_4$). The obtained $k_1$ values confirm the good response of the materials to degrade MB compared to the previous reports of heterogeneous Fenton and Fenton-like catalysts which are reported in the $k_1$ =0.0053 min⁻¹- 0.1455 min⁻¹ range. [63–68] In the case of the adjustment with a second order equation, similar trends are obtained, with $k_2$=0.0020(3) min⁻¹mg⁻¹L for the magnetite at room temperature, and $k_2$=0.0039(4) min⁻¹mg⁻¹L for the sample $Cu_{0.4}Fe_{2.6}O_4$ (MW4) measured at 90 °C.

The regression coefficients obtained from the fitting, $R^2$, are in the range $R^2$=0.82-0.991 for the pseudo-first order reaction, and $R^2$=0.95-0.99997 for the pseudo-second order reaction model, which signal that the kinetic is better described with the pseudo-second order reaction equation. This result suggests that the amount of MB molecules adsorbed on the catalyst depends on its initial concentration and determines the kinetic of the reaction.[9]

| Sample | | ST0 | | ST3 | | MW1 | | MW2 | | MW4 | |
|---|---|---|---|---|---|---|---|---|---|---|---|
| Temp. | | RT | 60°C | RT | 60°C | 60°C | 90°C | 60°C | 90°C | 60°C | 90°C |
| PFO | $k_1$ (min⁻¹) | 0.07(2) | 0.010(1) | 0.009(1) | 0.0032(2) | 0.0028(7) | 0.032(4) | 0.0044(5) | 0.065(9) | 0.009(1) | 0.10(2) |
| | $R^2$ | 0.85 | 0.97 | 0.94 | 0.991 | 0.82 | 0.96 | 0.97 | 0.94 | 0.96 | 0.91 |

| PSO | $k_2$ (min$^{-1}$ mg$^{-1}$L) | 0.0020(3) | 0.00011(1) | 0.00017(3) | 0.000034(2) | 0.000029(8) | 0.000047(4) | 0.00010(1) | 0.00049(2) | 0.00159(7) | 0.0039(4) |
|---|---|---|---|---|---|---|---|---|---|---|---|
| | $R^2$ | 0.95 | 0.9994 | 0.9990 | 0.99997 | 0.9995 | 0.998 | 0.9998 | 0.996 | 0.9993 | 0.97 |

**Table 2.** Reaction constants for methylene blue degradation: Data fitted using Pseudo-First-Order (PFO) and Pseudo-Second-Order (PSO) models. The table also includes the corresponding regression coefficients for each fit.

## 4. Conclusions

We have evaluated the catalytic efficiency of Cu doped magnetite and maghemite nanoparticles with average size of $\approx$11 nm, obtained from solvothermal and microwave-assisted methods, respectively. We found that the magnetite nanoparticles present an excellent performance for oxidizing the organic dye methylene blue at room temperature, producing up to a 90% of discoloration in 2 h. The catalytic activity was attributed to $Fe^{2+}$ centers that generate •OH in the heterogeneous Fenton reaction, which are the main species responsible for the MB oxidation. This degradation efficiency decreased notably with increasing Cu content. Consistently, the •OH species is systematically reduced when the Cu concentration increases. However, above room temperature, the magnetite losses its efficiency as nanocatalyst due to the surface oxidation and because of the low rate constant of the $Fe^{3+}/Fe^{2+}$ redox cycle in the Fenton reaction. In this condition the role of copper became relevant, with the samples with the highest percentage of copper being those that presented a better efficiency as catalysts. In fact, at 90°C the $Cu_xFe_{3-x}O_4$ with x=0.4 produce a 90% of MB discoloration in the first 30 minutes and a complete discoloration in less than 2 h.

These results show that the nanoparticle composition and the oxidation state of the surface active ions, determine the nanoparticle reactivity and the nature of the free radicals produced, providing a tool to engineering more efficient nanocatalysts. In particular, we have demonstrated that while magnetite nanocatalyst is effective to degrade dyes at room temperature, it is also highly unstable compound, easily prone to oxidation under normal conditions, and particularly at high temperatures. Instead, copper ferrite has an advantage over magnetite in terms of stability being a superior alternative for catalytic applications at higher temperature. Given that many industrial residual effluents are produced above room temperatures, as observed in textile and paper industries, leveraging this additional kinetic energy becomes interesting for enhancing the performance of nanocatalysts in decomposing organic contaminants. In summary, our research contributes valuable insights that could pave the way for the development of more efficient nanocatalysts, with practical applications in addressing environmental challenges associated with industrial effluents.

## 5. Declaration of interest

The authors declare that they have no known competing financial interests or personal relationships that could have appeared to influence the work reported in this paper.

## 6. CRediT authorship contribution statement

**Nahuel Nuñez:** Writing – Original Draft, Conceptualization, Investigation, Visualization. **Enio Lima Jr.:** Investigation. **Marcelo Vásquez Mansilla:** Investigation. **Gerardo F. Goya:** Funding acquisition, Project administration. **Álvaro Gallo-Cordova:** Investigation. **Maria del Puerto Morales:** Supervision, Resources. **Elin L. Winkler:** Writing – Review & Editing, Conceptualization, Supervision, Resources. The manuscript was written through contributions from all authors. All authors have given approval to the final version of the manuscript.

## 7. Acknowledgement

The authors thank to the Argentine government agency ANPCyT for providing financial support for this work through Grants No. PICT-2019-02059, as well as to UNCuyo for their support through Grant No. 06/C029-T1. Additionally, this research received partial funding from Project PDC2021-12109-I00 (MICRODIAL) MCIN/AEI/10.13039/501100011033 through the European Union "NextGenerationEU/PRTR". Furthermore, the authors acknowledge the support of Project H2020-MSCA-RISE-2020 (NESTOR) PROJECT Nº 101007629, funded by the EU-commission.

# Effect of the temperature and copper doping on the heterogeneous Fenton activity of $Cu_xFe_{3-x}O_4$ nanoparticles

Nahuel Nuñez[1,2,3*], Enio Lima Jr.[1,2], Marcelo Vásquez Mansilla[1,2], Gerardo F. Goya[4,5],

Álvaro Gallo-Cordova[6], María del Puerto Morales[6], Elin L. Winkler[1,2,3*]

[1] Resonancias Magnéticas, Gerencia de Física, Centro Atómico Bariloche, Av. Bustillo 9500, (8400) S. C. de Bariloche (RN), Argentina.

[2] Instituto de Nanociencia y Nanotecnología (CNEA-CONICET), Nodo Bariloche, Av. Bustillo 9500, (8400) S. C. de Bariloche (RN), Argentina.

[3] Instituto Balseiro, CNEA-UNCuyo, Av. Bustillo 9500, (8400) S. C. de Bariloche (RN), Argentina

[4] Dept. Física de la Materia Condensada, Universidad de Zaragoza, C/ Pedro Cerbuna 12, 50009, Zaragoza, Spain

[5] Instituto de Nanociencia y Materiales de Aragón, CSIC-Universidad de Zaragoza, C/ Mariano Esquillos S/N, 50018, Zaragoza, Spain

[6] Instituto de Ciencia de Materiales de Madrid, ICMM/CSIC, C/ Sor Juana Inés de la Cruz 3, 28049, Madrid, Spain

*Corresponding authors : nahuel.nunez@ib.edu.ar

**XPS characterization.** We performed XPS measurements of the Fe $2P_{3/2}$ absorption of solvothermal and microwave-assisted samples, which are presented in **Fig.S1** and **Fig.S2** respectively. The spectra obtained were fitted with two multiplets related to $Fe^{2+}$ and $Fe^{3+}$ respectively, as done by Grosvenor *et al.*[45], and the fitting parameters are presented in **TableS1**.

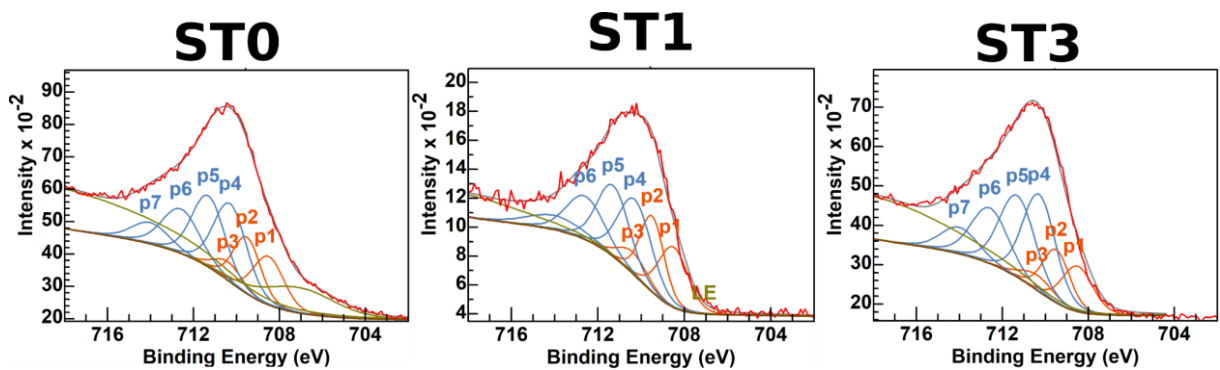

**Fig.S1:** XPS spectra and fit of the Fe $2P_{3/2}$ absorption in the solvothermal samples. Peaks p1, p2 and p3 are from $Fe^{2+}$ ion; peaks p4, p5, p6 and p7 are from $Fe^{3+}$ ion. The peaks position for each of the 3 peaks related to the **$Fe^{2+}$** and 4 peaks related to the **$Fe^{3+}$** were extracted from reference 45

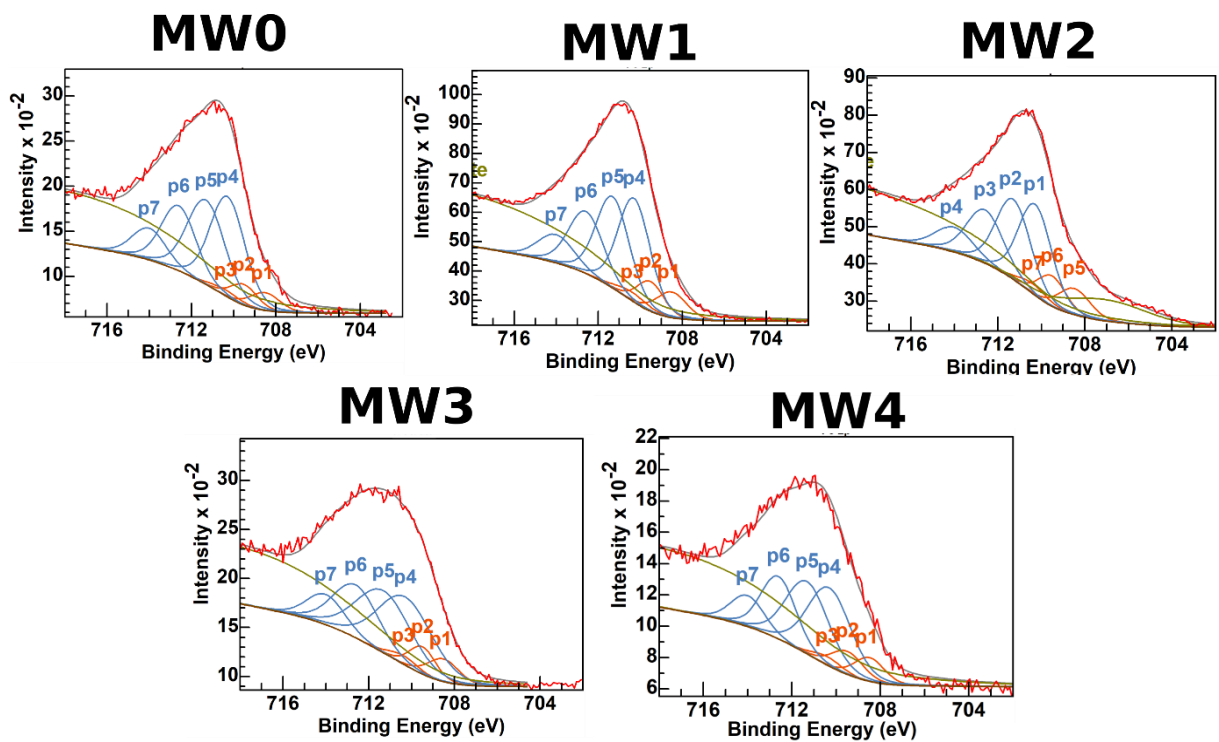

**Fig.S2:** XPS spectra and fit of the Fe $2P_{3/2}$ absorption in the microwave-assisted samples. Peaks p1, p2 and p3 are from $Fe^{2+}$ ion; peaks p4, p5, p6 and p7 are from $Fe^{3+}$ ion. The peaks position for each of the 3 peaks related to the $Fe^{2+}$ and 4 peaks related to the $Fe^{3+}$ were extracted from reference. [45]

| Identification | $Fe^{2+}$ | | | $Fe^{3+}$ | | | | Satellite | | |
|---|---|---|---|---|---|---|---|---|---|---|
| peak | p1 | p2 | p3 | p4 | p5 | p6 | p7 | LE | HE | |
| BE (eV) | 708.50 (2) | 709.50 (3) | 710.40 (1) | 710.30 (2) | 711.30 (2) | 712.60 (3) | 714.00 (4) | 707.1 (2) | 718.9 (5) | |
| Sample | Area | | | | | | | | | $Fe^{2+}$/ Fe ratio |
| ST0 | 583 | 600 | 209 | 984 | 895 | 636 | 305 | 662 | 3579 | 0.33 |
| ST1 | 125 | 129 | 45 | 218 | 199 | 141 | 67 | 0 | 2066 | 0.32 |
| ST3 | 406 | 419 | 146 | 954 | 868 | 617 | 295 | 0 | 1997 | 0.26 |
| MW0 | 106 | 110 | 38 | 689 | 627 | 446 | 213 | 0 | 2555 | 0.11 |
| MW1 | 334 | 344 | 120 | 1435 | 1306 | 928 | 444 | 0 | 5040 | 0.16 |
| MW2 | 270 | 278 | 97 | 996 | 906 | 644 | 308 | 595 | 3754 | 0.18 |
| MW3 | 85 | 87 | 30 | 540 | 491 | 349 | 167 | 0 | 2359 | 0.12 |
| MW4 | 78 | 81 | 28 | 351 | 319 | 227 | 108 | 0 | 1350 | 0.16 |

**Table S1:** Parameters used for the fitting of the XPS spectra. The peaks position for each of the 3 peaks related to the $Fe^{2+}$ and 4 peaks related to the $Fe^{3+}$ were extracted from reference [45]. The relationship between the area of the peaks of each multiplet was fixed. Peaks LE and HE corresponds to low and high energy satellites.

**Stability analysis.** We investigated the aging stability of selected catalysts under various conditions, as well as the structural changes in the materials after the catalytic reaction. Specifically, using EPR, we quantified the •OH free radicals production in the as made MW1 and MW4 samples, and again after four months of storage in air and water. These results are detailed in **Fig.S3**. Additionally, we examined the structural alterations of ST0 and MW4 samples post-catalysis. TEM images of samples ST0 and MW4 after 6 hours of continuous catalysis are presented in **Figs.S4-b** and **-d**, respectively. Correspondingly, **Figs.S4-a** and **-c** show a comparative analysis of the X-ray diffraction patterns before and after the catalytic reaction.

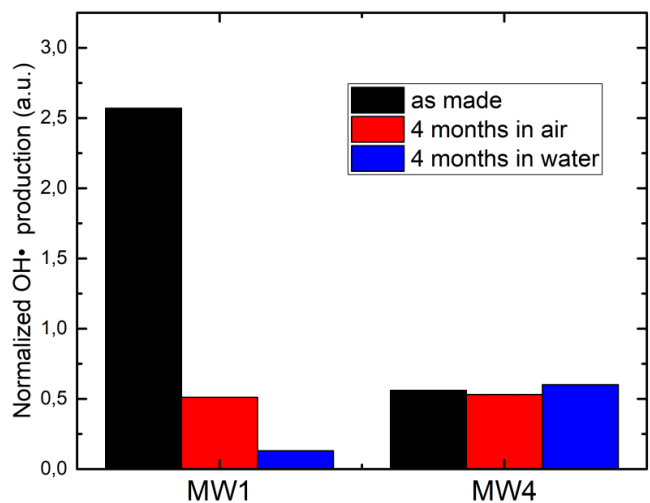

**Fig. S3:** Normalized •OH production, determined by EPR, of samples MW1 and MW4 in different conditions: as made (black) and after 4 months stored in air (red) and water (blue).

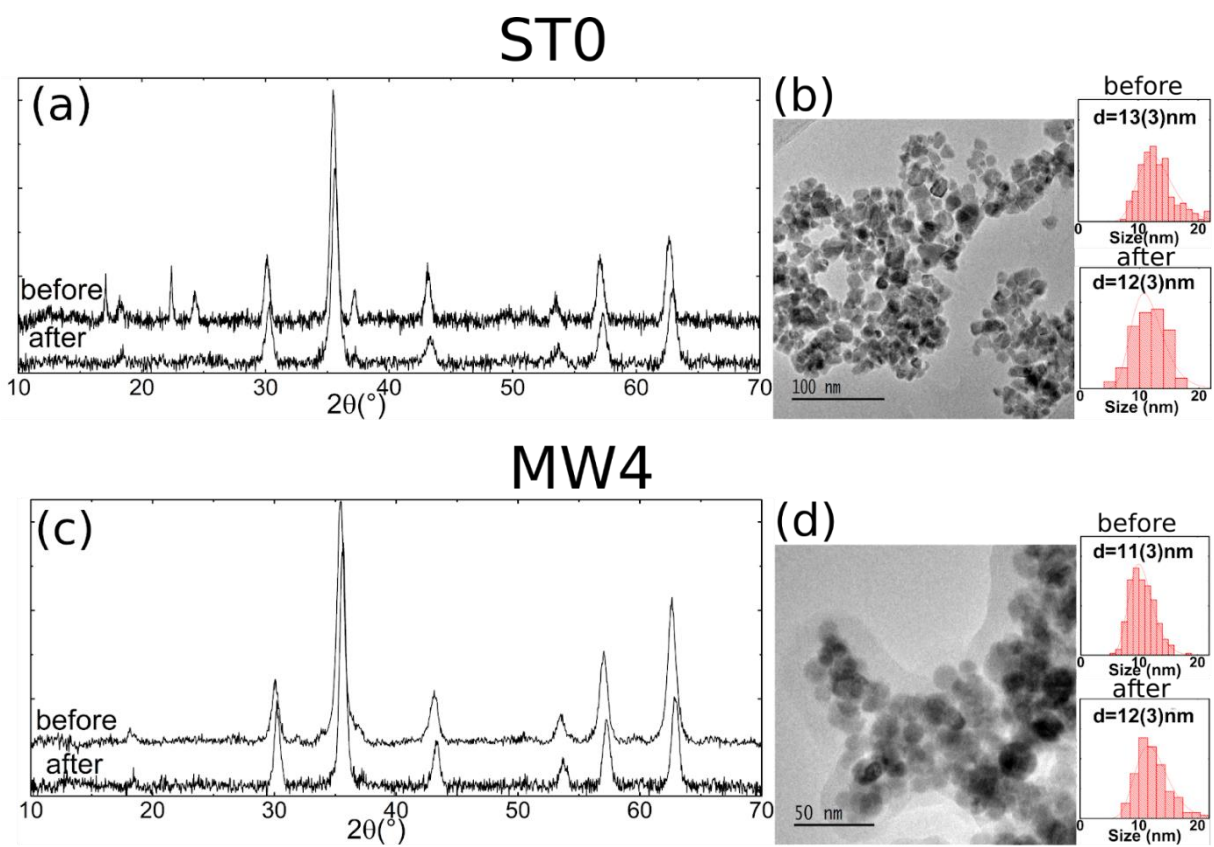

**Fig. S4: (a)** XRD patterns of ST0 before and after 6 hours of continuous catalysis. **(b)** TEM of ST0 post-catalysis with comparative size distributions before and after catalysis. **(c)** XRD patterns of MW4 before and after 6 hours of continuous catalysis. **(d)** TEM of MW4 post-catalysis with comparative size distributions before and after catalysis.

**Kinetic modelling.** From the methylene blue degradation curves depicted in **Fig.5**, different reaction kinetics models were explored. Specifically, the experimental data were fitted to pseudo-first-order $(\ln(\frac{C_0}{C}) = k_1 t)$ and pseudo-second-order $(\frac{C_0}{C} = k_2 C_0 t + 1)$ reaction models. **Fig.S5** illustrates the linearized pseudo-first-order fits for the various catalysts used under different experimental conditions. The linearized pseudo-second-order fits are shown in **Fig.S6. Table 2** presents the $k_1$ and $k_2$ constants derived from each model, along with their corresponding regression coefficients $R^2$.

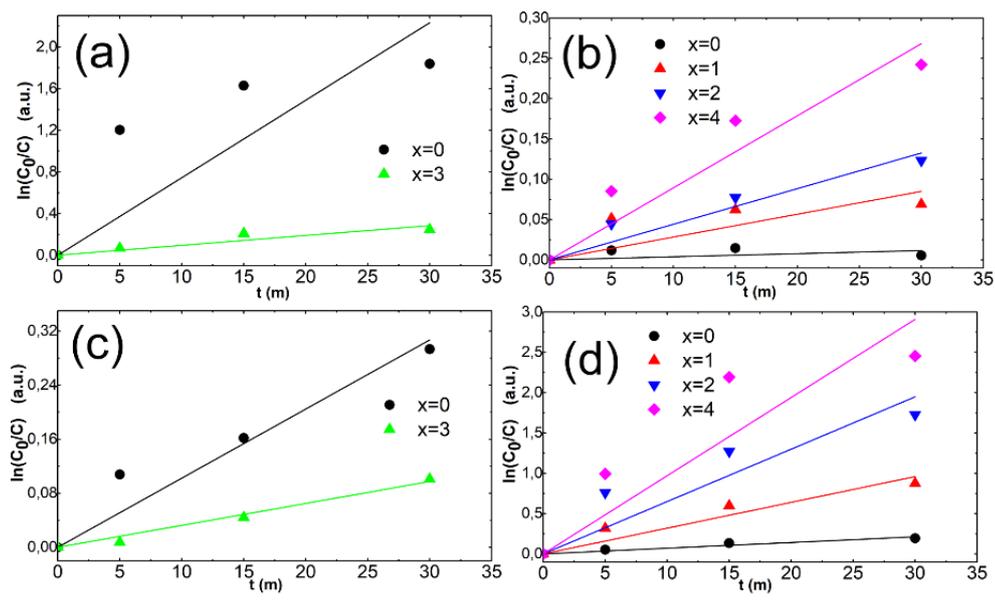

**Fig.S5:** Pseudo-First-Order kinetics for methylene blue degradation at different experimental conditions: **(a)** ST samples at room temperature (RT), **(b)** MW samples at 60°C, **(c)** ST samples at 60°C, and **(d)** MW samples at 90°C.

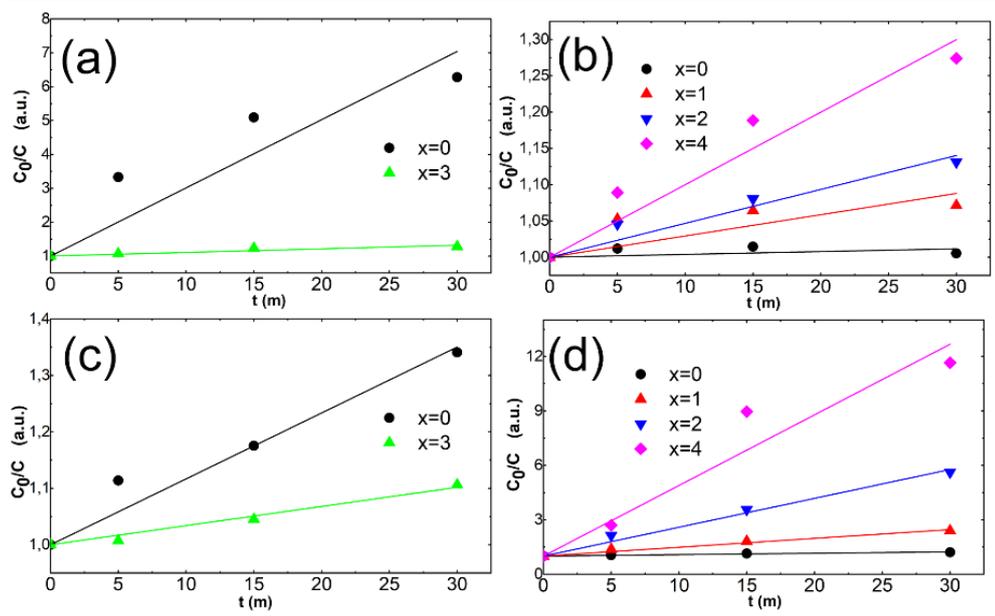

**Fig.S6:** Pseudo-Second-Order kinetics for methylene blue degradation at different experimental conditions: **(a)** ST samples at room temperature (RT), **(b)** MW samples at 60°C, **(c)** ST samples at 60°C, and **(d)** MW samples at 90°C.